\documentclass[useAMS,usenatbib]{mn2e}

\usepackage{url}
\usepackage{aas_macros}
\usepackage[english]{babel}
\usepackage[utf8x]{inputenc}
\usepackage{todonotes}
\usepackage{algorithmic}
\usepackage{algorithm}
\usepackage{units}
\usepackage{amsmath}
\usepackage{graphicx}
\def\Mpc{{\rm\thinspace Mpc}}
\def\Mdot{\hbox{$\dot M$}}

\def\Gpc{{\rm\thinspace Gpc}}
\newcommand\beq{\begin{equation}}
\newcommand\eeq{\end{equation}}
\newcommand\beqa{\begin{eqnarray}}
\newcommand\eeqa{\end{eqnarray}}
\def\Msun{\hbox{$\rm\thinspace M_{\odot}$}}
\def\msun{\hbox{$\rm\thinspace M_{\odot}$}}

\newcommand{\mpch}{\,h^{-1}\unit{mpc}}

\newcommand{\bluetides}{{\sc BlueTides}}

\title{The origin of most massive
  black holes at high-z: BLUETIDES and the next quasar frontier}

\author[Tiziana Di Matteo et al.] {Tiziana Di Matteo$^{1}$, Rupert A.C. Croft$^{1}$, Yu Feng$^{2}$, Dacen Waters$^{1}$, Stephen Wilkins$^{3}$ \\
$^1$\,McWilliams Center for Cosmology, Physics Dept., Carnegie Mellon
University, Pittsburgh PA, 15213, USA \\
$^2$\,Berkeley Center for Cosmological Physics, University of California at
Berkeley, Berkeley, CA 94720, USA \\
$^3$\,Astronomy Centre, Department of Physics and Astronomy, University of Sussex, Brighton, BN1 9QH, UK \\
}

\begin{document}
\maketitle

\begin{abstract}
  The growth of the most massive black holes in the early universe,
  consistent with the detection of highly luminous quasars at $z> 6$
  implies sustained, critical accretion of material to grow and power
  them.  Given a black hole seed scenario, it is still uncertain which
  conditions in the early Universe allow the fastest black hole
  growth. Large scale hydrodynamical cosmological simulations of
  structure formation allow us to explore the conditions conducive to
  the growth of the earliest supermassive black holes.  We use the
  cosmological hydrodynamic simulation \bluetides, which incorporates
  a variety of baryon physics in a $(400h^{-1} \mathrm{Mpc})^3$ volume
  with 0.7 trillion particles to follow the earliest phases of black
  hole critical growth. At $z=8$ the most massive black holes (a
  handful) approach masses of $10^{8} \Msun$ with the most massive
  (with $M_{\rm BH} = 4 \times 10^8 \Msun$) being found in an
  extremely compact spheroid-dominated host galaxy. Examining the
  large-scale environment of hosts, we find that the initial tidal
  field is more important than overdensity in setting the conditions
  for early BH growth. In regions of low tidal fields gas accretes
  'cold' onto the black hole and falls along thin, radial filaments
  straight into the center forming the most compact galaxies and most
  massive black holes at earliest times.  Regions of high tidal fields
  instead induce larger coherent angular momenta and influence the
  formation of the first population of massive compact disks. The
  extreme early growth depends on the early interplay of high gas
  densities and the tidal field that shapes the mode of accretion.
  Mergers play a minor role in the formation of the first generation,
  rare massive BHs.
\end{abstract}

\section{Introduction}

It is now well established that the properties of supermassive black
holes (SMBH) found at the centers of galaxies today are tightly
coupled to those of their hosts implying a strong link between black
hole and galaxy formation. The strongest direct constraint on the
high-redshift evolution of SMBHs comes from the observations of
luminous quasars at $z \sim 6$ in the Sloan Digital Sky Survey (SDSS)
\citep{Fan2006, Jiang2009} and more recently the highest redshift
quasar discovered at $z \sim 7$ \citep{Mortlock2011}. No quasars have
yet been observed at higher redshifts.  Although extremely rare (the
comoving space density of $z \sim 6$ quasars is roughly $n \sim$ a
few$ \Gpc^{-3}$) the inferred hole masses of these quasars are in
excess of $10^9~\Msun$, comparable to the masses of the most massive
black holes in the Universe today.  The origin of the seeds for these
black holes and the physical conditions that allow early growth to
supermassive black holes therefore remain a challenging problem.

In order to have had sufficient time to build up via gas accretion and
BH mergers (resulting from the hierarchical merging of their host
halos) the first 'seed' black holes must have appeared at early
epochs, $z > 10$.  The origin and nature of this seed population
remain uncertain. Two distinct populations of seed masses, in the
range of $100-10^5~\msun$ have been proposed: the small mass seeds are
usually thought to be the remnants of the first generation of PopIII
stars formed from metal-free gas at $z\sim 20-30$
\citep[e.g.][]{Bromm1999, Abel2000, Nakamura2001, Yoshida2003, Gao2006},
while the large seeds form by direct dynamical collapse in metal-free
galaxies \citep[][although see also Mayer et al. 2010 for direct
collapse into a massive blackhole in the metal-enriched
regime]{Koushiappas2004, Begelman2006}.

Growing the seeds to $~10^9 \Msun$ in less than a billion years
requires extremely large accretion rates - as mergers between black
holes are too rare and too inefficient for significant growth.  For a
black hole accreting at the critical Eddington accretion rate its
luminosity $L_{\rm Edd} = (4 \pi G c m_p)/ \sigma_{T} M_{BH} = \eta
\Mdot_{\rm Edd} c^2$ (where $G$, $c$, $m_{p}$ and $\sigma_{T}$ are the
gravitational constant, speed of light, proton mass and Thomson cross
section, and $\eta$ is the standard accretion efficiency) implies
exponential growth at the characteristic Eddington timescales, $t_{\rm
  Edd} = 450 \eta / (1- \eta)$ Myr, such that $M_{\rm BH} = M_{\rm
  seed} e^{t/t_{\rm Edd}}$. For a seed mass ranging from $M_{\rm seed}
\sim 100-10^5 \Msun$ this requires $10-17$ e-foldings to reach $M_{\rm
  BH} \sim 10^9 \Msun$. The crucial question (for any given seed
model) is then where (if at all, or in which kind of halos) and how
such vigorous accretion can be sustained at these early times.

As bright quasars are likely to occur in extremely rare high-density
peaks in the early universe, large computational volumes are needed to
study them. In \citet{DiMatteo2012} we used the MassiveBlack
simulation, a cosmological hydrodynamic simulation covering a volume
$(0.75 \Gpc)^3$ appropriate for studying the rare first quasars to
show that steady high density cold gas flows responsible for
assembling the first galaxies produce the high gas densities that lead
to sustained critical accretion rates and hence rapid growth
commensurate with the existence of $\sim 10^9 \Msun$ black holes as
early as $z\sim 7$.  Here we use our newest, higher resolution
cosmological hydrodynamic simulation, \bluetides \citep{Feng2016,
  Feng2015, Waters2016a, Waters2016b} (covering a volume of $[0.5
\Gpc]^{3}$) which includes improved prescriptions for star formation,
as well as black hole accretion and associated feedback processes. We
focus on earlier times, searching for the conditions that allow rapid
growth of the massive black holes beyond currently observed redshifts
($z >7$). This also allows us to make predictions for the most massive
black holes to be discovered in the near future. Large volume
simulations such as {\it MassiveBlack} \citep{DiMatteo2012} and
associated high-resolutions zooms \citep{Feng2014} already showed that
the existence of the $z=6$ SDSS quasar population, is consistent with
our standard structure formation models.  Crucially, the \bluetides
simulation with its increased resolution, allows us to study reliably
the formation of massive black holes at even earlier times.  It
therefore provides a unique framework to study the formation of the
next frontier of first quasars.

\begin{figure}
\hbox{
\includegraphics[width=\columnwidth]{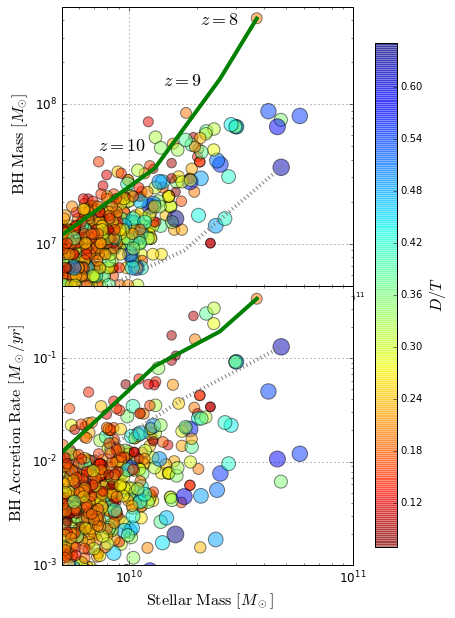}}
\caption{Top panel: Black hole mass versus host galaxy stellar mass at
  $z=8$.  Colors indicate the disk to total (bulge plus disk) ratio
  for the stellar component, so that blue colors indicate disks and
  yellow/orange spheroid dominated hosts. The most massive black hole
  at $z=8$ in BlueTides has a mass of $M_{\rm BH} = 4\times 10^{8}
  \Msun$ and resides in a spheroidal host. There are about half a
  dozen black holes with masses close to $10^{8} \Msun$. The green and
  dashed track show the redshift evolution of the most massive BH and
  a massive disk in this plane.}
\label{fig:mbhmstar}
\end{figure}

\section{BlueTides Simulation}
The \bluetides\ simulation (see Feng et al. 2015, 2016) for a full
description) was carried out using the Smoothed Particle Hydrodynamics
code {\sc MP-Gadget} with $2\,\times\, 7040^{3}$ particles on the Blue
Waters system at the National Centre for Supercomputing
Applications. The simulation evolved a $(400/h)^{3}\,{\rm cMpc^3}$
cube to $z=8$ by which time it contained approximately 200 million
objects (of which 160,000 have stellar masses greater than
$10^{8}\,{\rm M_{\odot}}$). At $z=12$ the number of objects identified
falls to around 20 million, and only around 30000 have stellar masses
greater than $10^{10}\,{\rm M_{\odot}}$. The galaxy stellar mass and
rest-frame UV luminosity functions predicted by the simulation
\citep{Feng2015, Feng2016, Waters2016a, Wilkins2016b, Wilkins2016c}
match observational constraints available at $z= 8, 9, 10$
\citep[e.g][]{Bouwens2015, Song2015}. The large volume of BlueTides,
almost 200 times larger than either the Illustris,
\citep{Vogelsberger2014} or EAGLE, \citep{Schaye2015} simulations,
makes it ideally suited to an investigation of rare objects at
high-redshift. The high resolution (comparable to Illustris) also
allows the study of the formation of the first (massive galaxies) and
details of the structure and gas inflows inside galaxies.  BlueTides
was run assuming the Wilkinson Microwave Anisotropy Probe nine year
data release \citep{Hinshaw2013}. The simulation is described in more
detail in \citep{Feng2015, Feng2016} while the physical and observed
properties of the galaxy population and predictions for the black hole
AGN populations and luminosity functions are described in detail in
\citep{Feng2015, Feng2016, Wilkins2016a, Wilkins2016b, Waters2016a,
  Waters2016b}.  In producing galaxy photometry we assume a Chabrier
(2003) initial mass function and utilise the Pegase.2 stellar
population synthesis model \citep{Fioc1997}.

A variety of sub-grid physical processes were implemented to study
their effects on galaxy formation:
\begin{itemize}
\item star formation based on a multi-phase star formation model
  \citep{Springel2003a} with modifications following
  \citet{Vogelsberger2013}
\item gas cooling through radiative processes \citep{Katz1996} and
  metal cooling \citep{Vogelsberger2014}
\item formation of molecular hydrogen and its effects on star
  formation \citep{Krumholz2011}
\item type II supernovae wind feedback (the model used in
  Illustris\citep{Nelson2015})
\item Black growth and AGN feedback using the same model as in
  MassiveBlack I and II \citep{DiMatteo2005}

\item a model of ``patchy'' reionization \citep{Battaglia2013}
  yielding a mean reionization redshift $z\approx10$
  \citep{Hinshaw2013}, and incorporating the UV background estimated
  by \citet{Faucher2009}

\end{itemize}

In \citet{Feng2015} we performed a standard kinematic decomposition of
the stellar component in the galaxies to determine the fraction of
stars in each galaxy which are on planar circular orbits, and which
are associated with a bulge. We note that this technique is not
sensitive to the clumpiness of disks. This provides kinematically
classified disks and bulges and a Disk to Total (D/T) ratio for our
galaxy sample above a mass of $10^{10}\msun$ (which contains a
sufficient number of stellar particles for this decomposition to be
carried out).

Here we also add a calculation of the large scale tidal field. The
tidal field can be characterized by three eigenvalues of the local
tidal tensor, $t1$, $t2$ and $t3$ (by definition $ t1 > t2> t3$) and
the corresponding eigenvectors ${\bf t1}$, ${\bf t2}$ and ${\bf
  t3}$. The three eigenvalues satisfy $t1 + t2 + t3 \equiv 0$ so that
$t1$ is always positive and $t3$ negative. Thus the tidal field
stretches material along $t1$ and compresses material along $t3$. By
using $t1$ as the indicator of the local tidal field strength we are
following standard usage.  We define the tidal field as the trace-less
part of the strain tensor
\begin{equation}
T_{ij} = S_{ij} - \frac{1}{3} \sum_i S_{ii},
\end{equation}
where the strain tensor $S_{ij}$ is the second derivative of the
potential $\phi$,
\begin{equation}
S_{ij} = \nabla_i \nabla_j \phi.
\end{equation}

The large scale strain tensor is calculated in Fourier space, following
the procedures in \cite{2008ApJ...687...12D},
\begin{equation}
\hat{S}_{ij} = \frac{k^2}{k_i k_j} \hat{\delta}.
\end{equation}

We assign the a uniform 1/1000th subsample of the BlueTides particles
to the a mesh of $512$ grid points per side by a Cloud-in-Cell window
\citep{1981csup.book.....H}. The density field is smoothed by a
Gaussian window $\exp -0.5 k^2 s^2$ with a width of $s = 1,2,5\,h^{-
  1}\mathrm{Mpc}$. The strain tensor value on the mesh is re-sampled
at the location of the central black hole particles by a Cloud-in-Cell
window.  We note that our definition is different from
\cite{2011MNRAS.413.1973W}, where the calculation is performed in
configuration space from the nearest neighbour galaxies.

\section{Results}

\subsection{The early assembly of most massive black holes}
We turn directly to the most massive black hole population in the
galaxies in Bluetides. The left panel of Figure~\ref{fig:mbhmstar}
shows black hole mass versus stellar mass of the the host galaxies at
$z=8$. The colours of the points indicate the D/T ratio for the
stellar component of each galaxy. The most massive black hole in
BlueTides has a mass of $4 \times 10^{8} \Msun$ at
$z=8$. Interestingly, the most extreme BH does not appear to be hosted
by the largest galaxy, in terms of stellar mass, and it has a
signifcantly more dominant spheroidal component ($D/T$ only 0.3) than
most of the more massive galaxies \citet{Feng2015}. The most massive
black hole is (positively) offset from the main $M_{BH}-M_{*}$
relation defined by the rest of the population indicating a relatively
faster growth of the black hole compared to the stellar host, as
expected. There are a handful of objects with $M_{\rm BH}$ approaching
$10^8 \Msun$, and only a couple of those in the most massive galaxies
with $D/T > 0.5$. A large fraction of the massive black hole
population is hosted in intermediate stellar mass hosts with
relatively dominant spheroidal components. Examining the left panel of
Figure~\ref{fig:mbhmstar} we can see by that both the orange points
(spheroids) and blue points (disks) follow a relationship between
stellar mass and black hole mass, but that the locus of the spheroidal
dominated galaxies is above (higher BH mass) the disks.

Looking at the black hole accretion rates (right panel of
Figure~\ref{fig:mbhmstar}) we can see that the largest black hole
accretion rates are also typically found in the black holes hosted in
the more spheroidal galaxies. The locus of black hole accretion rate
versus stellar mass for the spheroids is also displaced towards higher
black hole mass.  In summary therefore, for a given stellar mass, the
majority of the most massive black holes seen at $z=8$, and those
accreting at the largest rates, appear to be hosted by (extremely
compact) spheroidal dominated galaxies. In \citet{Feng2015} we
discussed the population of massive galaxies with significant disk
components.

\subsection{Examples of high-z massive black holes}
\subsubsection{Environments}
\begin{figure*}
\vbox{
\includegraphics[width=2\columnwidth]{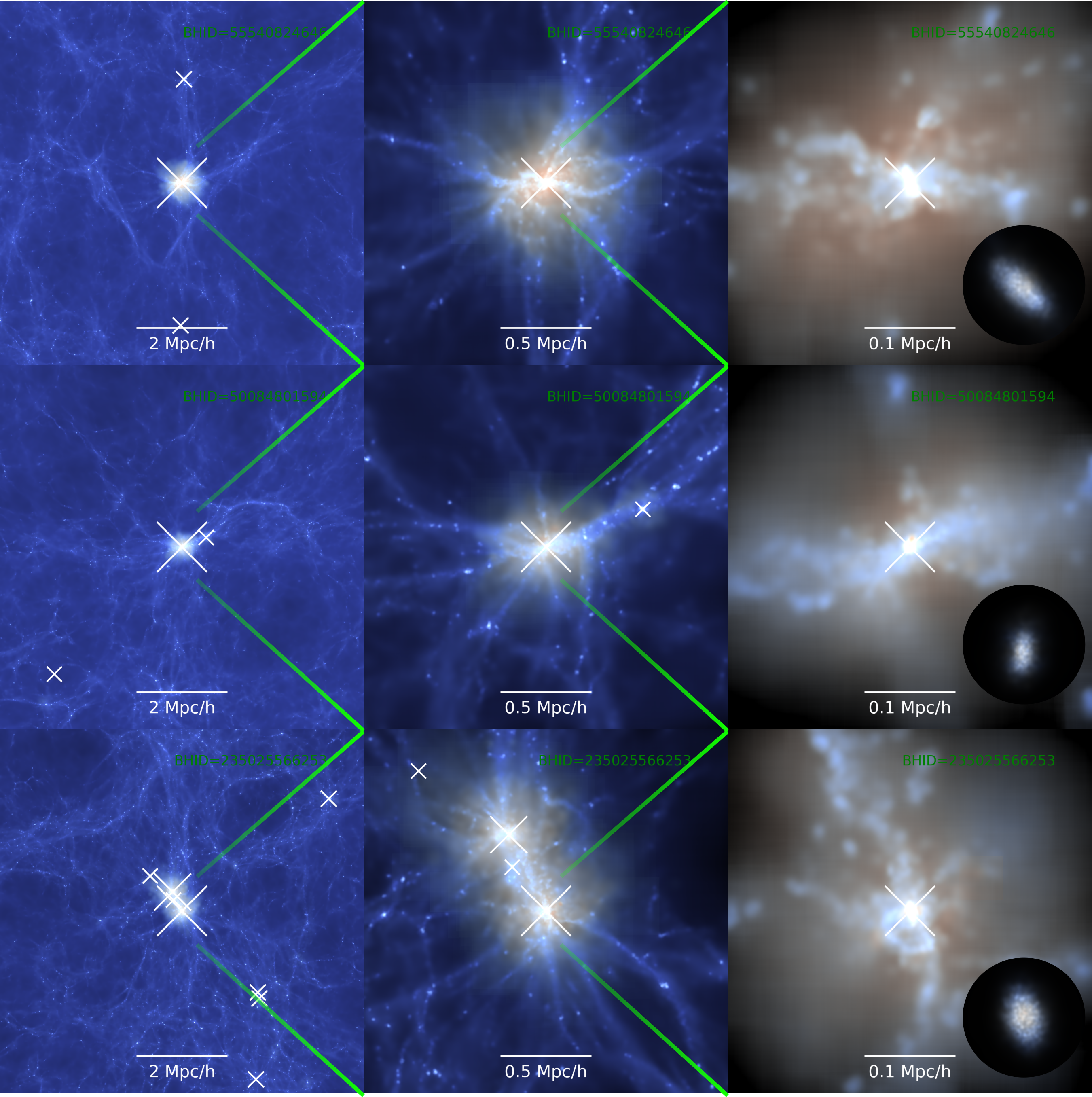}
}

\caption{The environments of the most massive $z=8$ black holes in
  \bluetides. The images show the large scale gas density around the
  black holes, zooming into regions of width 6 $\mpch$ (left panels) to 3
  $\mpch$ (center) and 0.5 $\mpch$ (right). The gas density field is
  color coded by temperature (blue to red indicating cold to hot
  respectively). The final zoom on the top right column show the
  stellar density for the host galaxy. The positions of the black holes
  in all images are indicated with white crosses the sizes of which are
  proportional to black hole mass.}
\label{fig:BHhosts}
\end{figure*}

\begin{figure*}
\vbox{
\includegraphics[width=2\columnwidth]{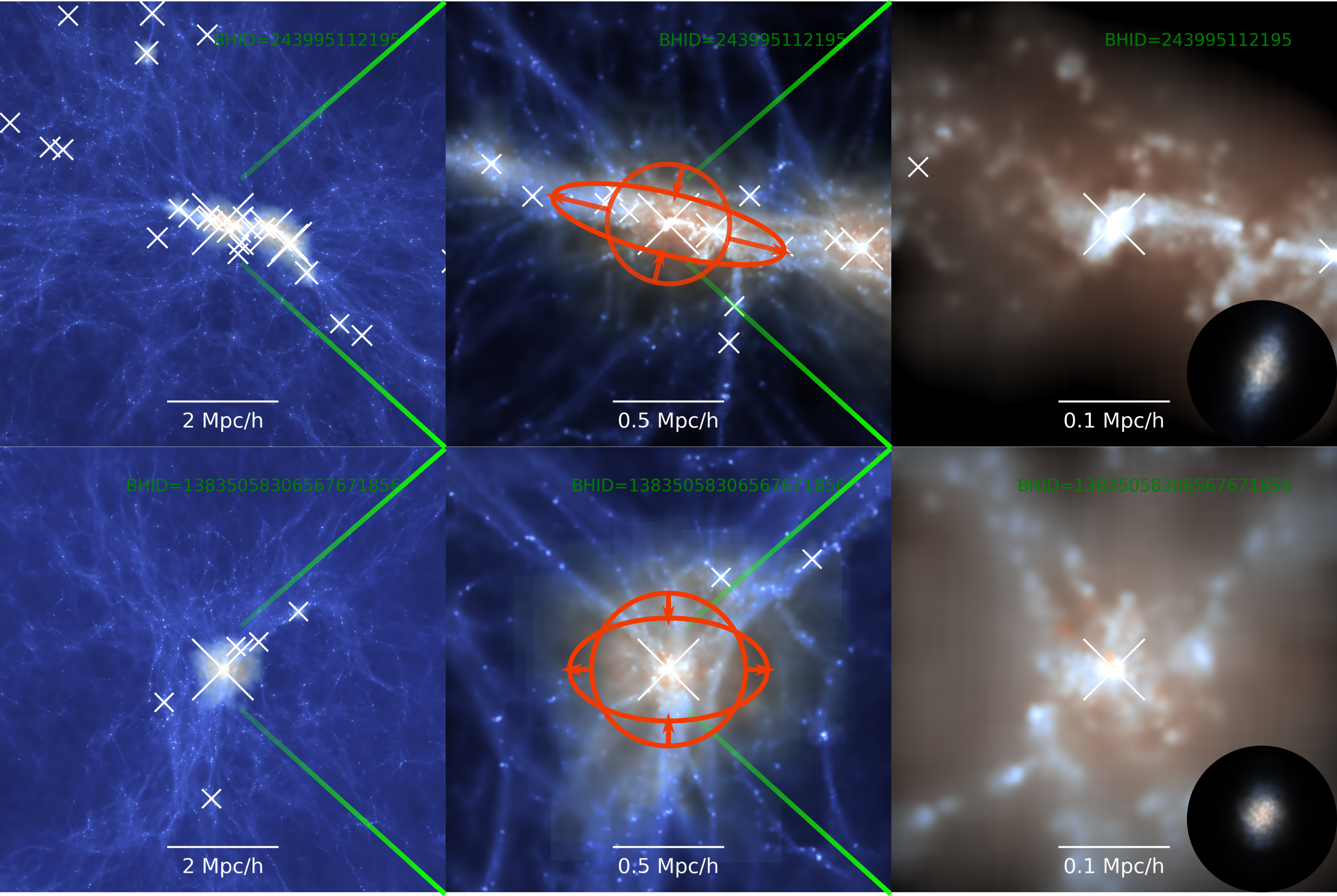}}
\caption{Same as Figure~\ref{fig:BHhosts}. The most massive disk
  galaxy environment (top row) compared to the most massive black hole
  and host galaxy (bottom row). The orange ellipses and arrow
  illustrate the tidal field $t1$ and $t3$ in the two cases. Tidal
  fiels stretches material along $t1$ and compresses material along
  $t3$.  Strong fields, large $t1$ is found in the large scale
  filament in the environment of the disk galaxy and small $t1$ and
  associated thin, 'cold' filaments are characteristic of the
  enviroments of the most massive BHs.}
\label{fig:BHhosts_disk}
\end{figure*}

To illustrate the environments and corresponding host galaxy
morphologies of the most massive black holes, we show in
Figure~\ref{fig:BHhosts} images of 3 regions containing the most
massive black holes in the simulation at $z=8$. In Figure
~\ref{fig:BHhosts_disk} we compare the environment and host of the
most massive black hole in \bluetides (with $M_{\fm BH} 4 \times
10^8\Msun$) with that of one of the massive disks, as discussed in
\citet{Feng2015}. For both figures, the images show the projected gas
density after zooming into scales 6 $\mpch$, 3 $\mpch$ and 0.5 $\mpch$
on a side (for left, center and right image in each set
respectively). The final inset in the right panel shows the central
galaxy host (adaptively smoothing the star particle distribution and
color coding according to SFR). The positions of the black holes in
all images are indicated with white crosses the sizes of which are
proportional to black hole mass. It is interesting to note that only
one of these massive black holes is close to undergoing a major
merger, and all other massive black holes appear to be associated with
isolated high density regions. It is also evident from these images
that the most massive black holes strongly favour quasi-spherically
collapsed regions of the density field and in turn that such
environments are more conducive to the formation to spheroid dominated
systems rather than disk dominated stellar hosts.  The massive
disk-dominated galaxies also host black holes but generally the black
hole mass is not amongst the most massive at these epochs. We
quantify the origin of the massive galaxy morphology and related black
hole growth in Section~\ref{Origin}.


\begin{table}
\centering
\begin{tabular}{||c c c c c ||} 
 \hline
 Property & BH1 & BH2 & BH3 & BH4 \\  
 \hline\hline
$M_{BH}(\times 10^{8} \Msun)$ &  4.1 & 1.5 & 0.9 & 0.8  \\
$M_{200} (\times 10^{10} \Msun) $ & 42 & 33   &44  &92 \\
$M_{\rm star}(\times 10^{10} \Msun) $  & 3.7 & 2.5  &4.1  &5.7  \\
$M_{\rm gas}(\times 10^{10} \Msun) $   & 4.0  & 3.9  & 4.9  & 13 \\
$f_{gas}[M_{\rm gas}/(M_{\rm gas} + M_{\rm star})]$ &0.5 & 0.6 & 0.6 & 0.7\\
$D/T$  & 0.2 & 0.1 & 0.5 & 0.5 \\
$t1$ (tidal field) &0.29&0.23& 0.32& 0.08\\
\hline
\end{tabular}
\caption{Basic properties of the four most massive black holes and their
hosts at $z=8$. Images of the environments and host galaxies of these objects 
are shown in Fig.~\ref{fig:BHhosts}.}
\label{table:BHhost_properties}
\end{table}

\subsubsection{Evolution}
\begin{figure}
\includegraphics[width=\columnwidth]{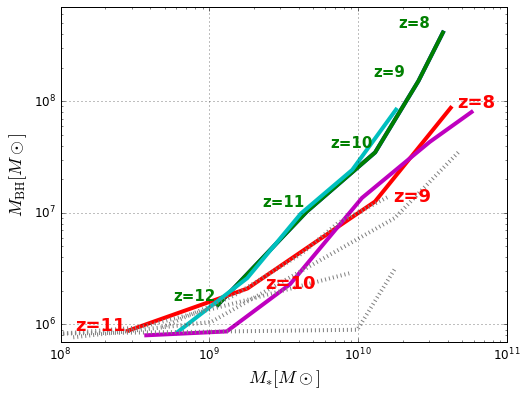}
\caption{The Coevolution of blackhole mass and stellar mass in the
  four most massive black holes (solid lines) compared to four
  of the most massive disks (dashed lines). Black hole growth is
  faster in the the spheroid-dominated objects compared to the growth
  in disk-dominated galaxies.}
\label{fig:mbhmstarevol}
\end{figure}

In this section we study the evolution of a number of properties for
the most massive black holes and compare them to those of some of the
most massive stellar systems. Figure~\ref{fig:mbhmstarevol} shows the
co-evolution from $z=12$ to $z=8$ of black hole mass and stellar mass
for the sample of the four most massive black holes at $z=8$ shown in
Figure~\ref{fig:BHhosts} and Figure~\ref{fig:BHhosts_disk} (solid lines).
Each line represent a single black hole as it moves from the lower
left region and ends up in the top right.  The x-axis is the
corresponding stellar mass of its host galaxy.  For comparison, the
dashed lines show the tracks for three other black holes which reside
in some disk-dominated hosts (one of which is shown in the top panel
of Fig.~\ref{fig:BHhosts_disk}). The tracks of black hole mass versus
stellar mass are steeper for the most massive black holes indicating a
relatively faster black hole growth than host stellar mass growth
comparedfor example to the disk-dominated systems. The latter reside
in the most massive stellar systems (dashed lines) which instead tend
to grow in stellar mass relatively faster than in black hole mass. At
this early time, a very rapid phase of black hole growth is required
in order to explain the extremely massive rare objects that are
seen. This rapid growth is consistent with the black hole leading
galactic growth through early assembly.

\begin{figure*}
\hbox{
\includegraphics[width=\columnwidth]{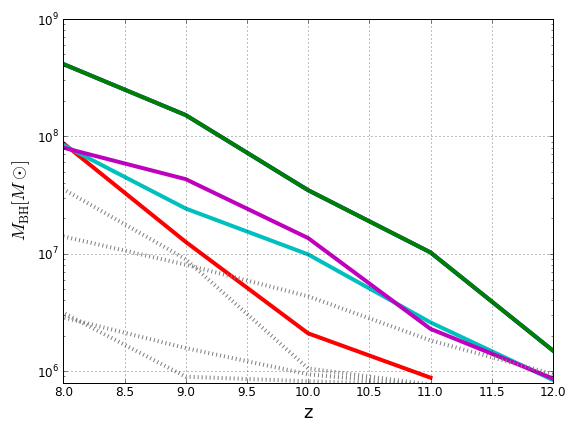}
\includegraphics[width=\columnwidth]{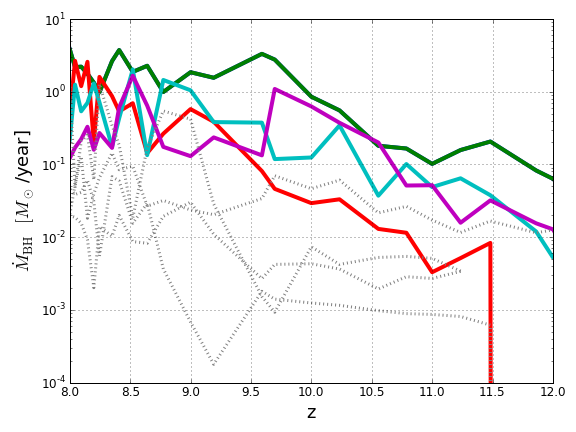}}
\caption{The evolution of blackhole and accretion rate.  Solid lines:
  blackhole mass and corresponding accretion rate for the most massive
  blackholes at $z=8$.  Dashed lines: a sample of the black
  holes in the most disk-dominated galaxies at $z=8$.}
\label{fig:mbhevol}
\end{figure*}

Figure~\ref{fig:mbhevol} tracks directly the evolution of the black
hole mass and accretion rate for the four most massive black holes
(solid lines) and those of the most massive, disky stellar systems. As
expected, the black holes that build up the largest masses accrete
close to their critical Eddington accretion rate virtually since the
time they were seeded.  This can be seen as the curves are close to
flat for most of the cosmic time. Note however that the black holes
are not exclusively following the Eddington rate but show some
variability, with peaks and troughs while remaining close to that
value.  These are a result of the complex interplay of AGN feedback in
the gas. While feedback is unable to quench and completely heat and
remove the accreting gas the inflows are disturbed and accretion rate
has significant variations particularly toward lower redshifts ($z\sim
8$). In the left panel of Figure~\ref{fig:mbhevol} we can see that at
this epoch black hole mass is still growing steeply, with no sign that
it has saturated at its final value. This is consistent with
observations of SDSS quasars which have implied masses close to
$10^{9} \Msun$ at $z \sim 6$ \citep[][see also]{DiMatteo2012}.  From the
right panel of Figure~\ref{fig:mbhevol} we can see that the BH
accretion rates in disk-dominated objects are about an order of
magnitude lower than in the most massive black hole sample.

\begin{figure*}
\hbox{
\includegraphics[width=\columnwidth]{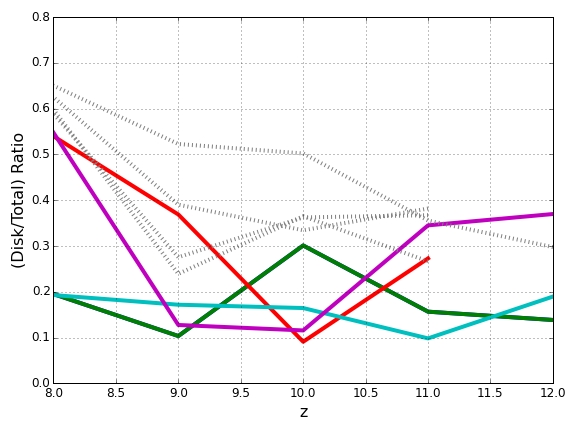}
\includegraphics[width=\columnwidth]{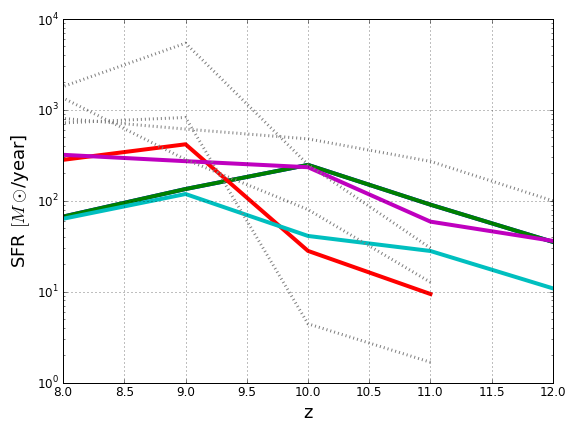}}
\caption{The D/T ratio for the host galaxies for black holes in 
Fig.~\ref{fig:mbhevol}. The solid lines and dashed lines
correspond to those in 
Fig.~\ref{fig:mbhevol}.}
\label{fig:mbhhostevol}
\end{figure*}

In Figure~\ref{fig:mbhhostevol} we turn to the evolution of the
associated D/T ratio and SFR of the host galaxies of the massive black
holes and massive disks discussed in Figure~\ref{fig:mbhevol}.

We have seen (Fig.~\ref{fig:mbhmstar} and Fig.~\ref{fig:BHhosts} that
the most massive black holes are typically associated with
spheroid-dominated hosts at $z=8$. Figure~\ref{fig:mbhhostevol} shows
that indeed the D/T ratios are typically small for the hosts of
massive black holes and in particular that D/T generally stays small
with redshift (at least over the redshift range we can access with
BlueTides). It is also interesting to note that while the typical
SFRs are pretty high at the highest redshifts for the massive black
holes in the spheroid-dominated hosts, they do not increase by much at
low redshifts. For the disk-dominated massive hosts on the other hand,
the star formation rates increase rapidly, reaching up to several
$1000 \Msun$/yr, up to factors 50 higher than in the massive black
hole hosts. It appears therefore that the presence and fast growth of
a massive black hole in these early compact spheroids is able to
quench (through its associated feedback) the SFR in the massive black
hole hosts. This somewhat evident also Fig.~\ref{fig:BHhosts} where
the stellar population in the central region of the spheroid is rather
old (shown by the the ref colors). We note however even at these
high-z gas rich environments the SFR are still a few hundred
$\Msun/yr$.

\begin{table*}
\centering
\begin{tabular}{||c c c c c ||} 
 \hline
 Property & $a$ & $b$ & $r_s$ & $\Delta$ \\  
 \hline\hline

Galaxy D/T & $9.75 \times 10^{-2}$ & 3.6 & $8.7 \times 10^{-2}$ & $1.4 \times 10^{-38}$ \\
BH Eddington & 0.22 & $6.12 \times 10^{-2}$ & $-4.94 \times 10^{-2}$ & $1.78\times 10^{-13}$ \\
\hline
\end{tabular}
\caption{Parameters $a$, and $b$ for a linear fit in the correlations between
  $D/T$ ratio and BH Eddington accretion rate versus tidal field strength $t1$ shown in Fig.~\ref{fig:tidalfield}. The spearman rank correlation coefficient $r_s$ and $\Delta$ the two sided
  significance of its deviation from zero. Here small values indicate more significant correlations.} 
\label{table:spearman}
\end{table*}

\begin{figure*}
\vspace{-1.5cm}
\hbox{
\hspace{-1.6cm}
\includegraphics[width=1.6\columnwidth]{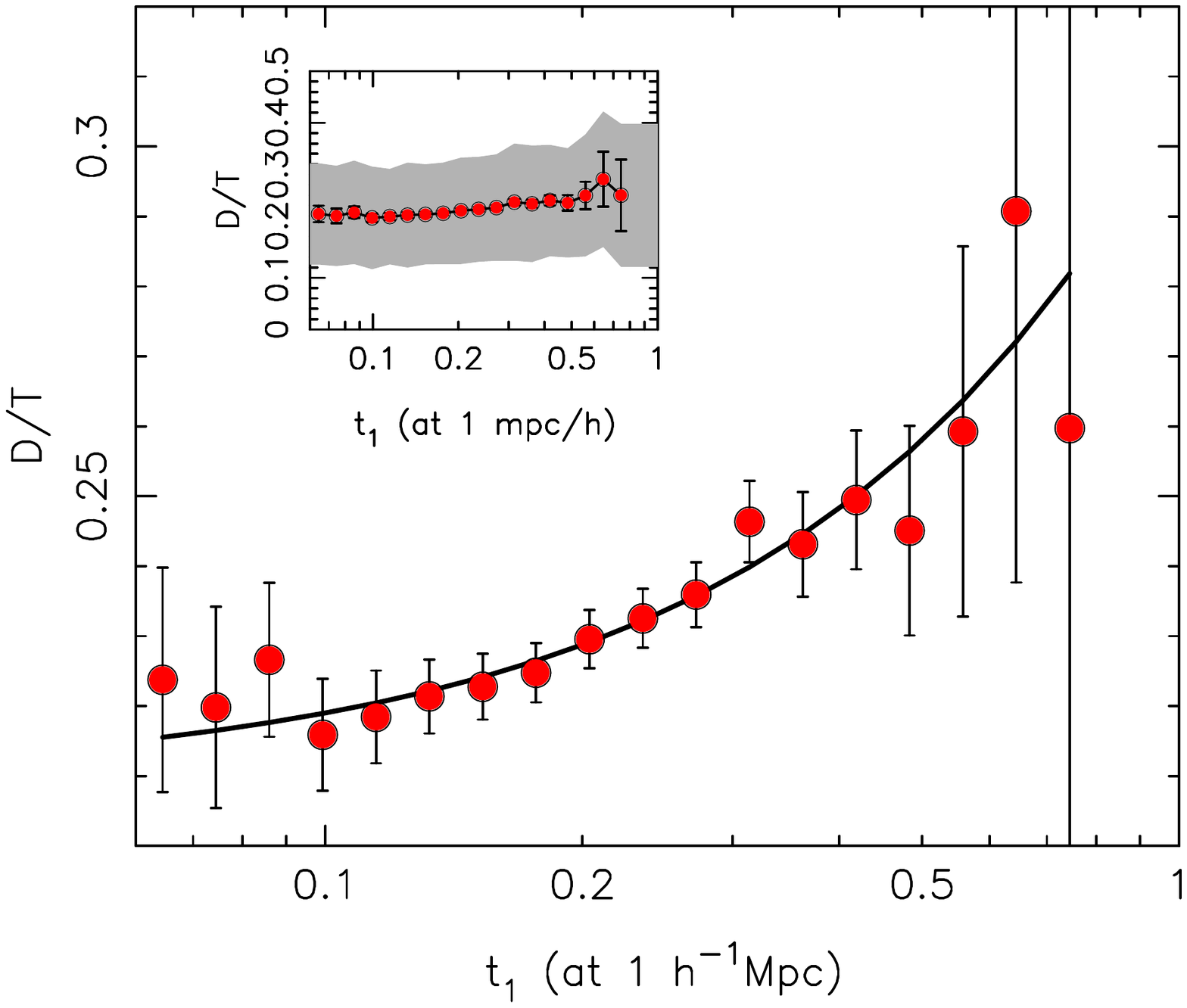}
\hspace{-4.1cm}
\includegraphics[width=1.6\columnwidth]{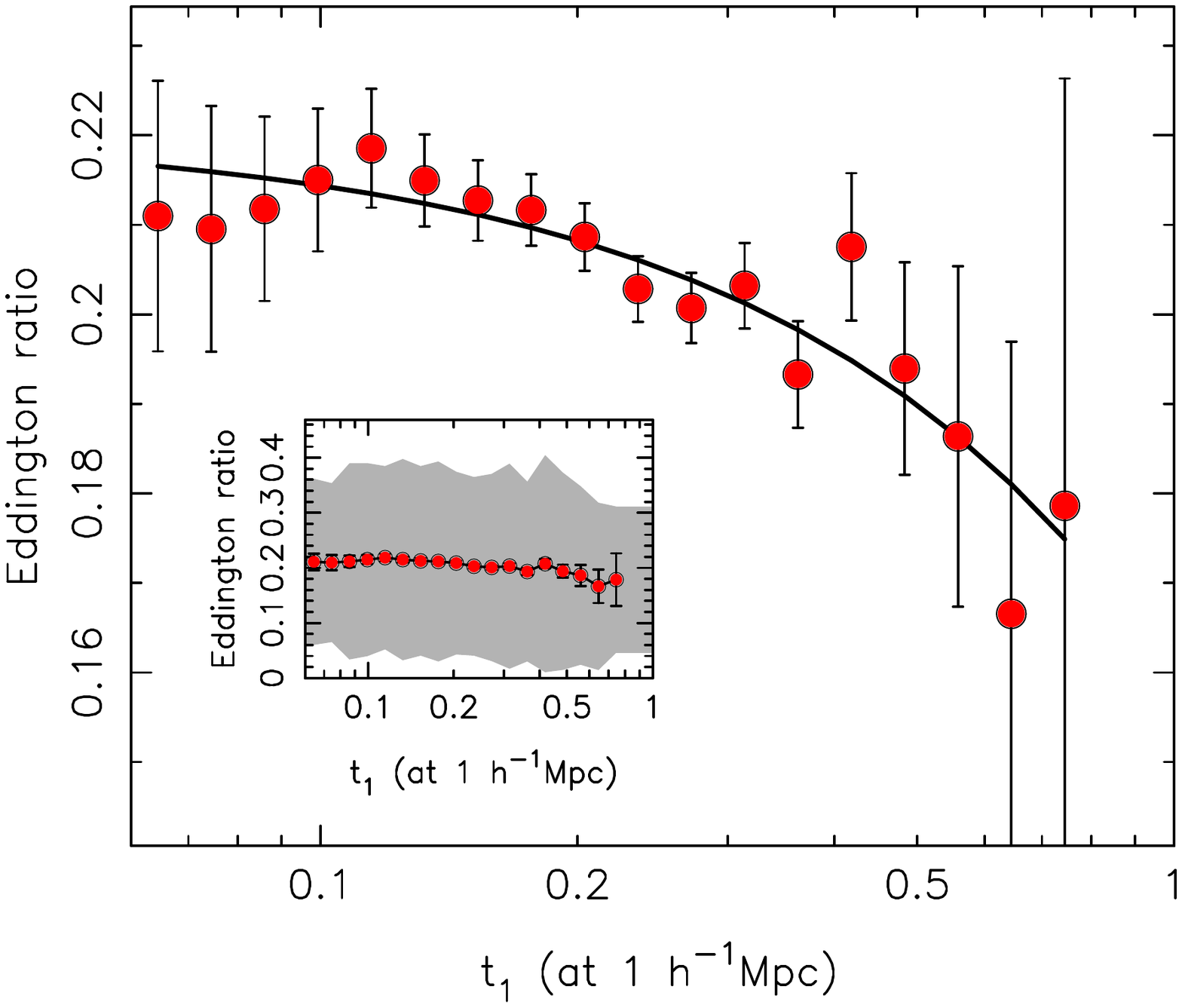}}
\vspace{-1cm}
\caption{Left: The disk-to-total ratio for  \bluetides BH host galaxies versus
  $t_{1}$, the tidal field strength. Right: the Eddington accretion rate
  of BHs versus $t_{1}$. The inset panels in each case show the same data, but
with an increased y-axis range. The points with error bars show the mean
values of D/T and Eddington ratio in bins of $t_{1}$. The error bar is the
error on the mean, and the shaded regions indicate the range which encloses 
68\% of the galaxies in each bin. The smooth curves are linear fits 
of D/T and Eddington ratio to $t_{1}$, with the fit parameters given in 
Table \ref{table:spearman}}
\label{fig:tidalfield}
\end{figure*}

\subsection{The origin of the fastest black hole growth: tidal field
strength}
\label{Origin}

Our results so far suggest that the growth of the most massive, early
black holes in BlueTides is linked to mechanisms operating on and
related to the formation of their hosts and not to the occurrence of
major mergers. It also appears that the galaxy morphology at these
epochs is related to black hole growth. Whether a galaxy is disk-like
or spheroidal is imprinted early on and not the result of a sudden
accretion/merger driven transformation. These considerations and the
visual differences in the large-scale distribution of surrounding
matter (illustrated by Figure~\ref{fig:BHhosts_disk}) in regions
surrounding host galaxies with rapid black hole growth versus regions
inducive of early disk formation suggest that net angular momentum and
the impact of tidal field strength in these regions may play a
role. In particular, we see that the large-scale distribution of the
gas around the massive black holes is typically composed of thin,
distinct filaments which do not interact and which lead to fast 'cold'
gas accretion to prevail forming a spheroid host. The formation of
spheroids in these environments has been seen in earlier work
\citep{Sales2012}. In that work it was also suggested that galaxy
morphology was related to spin alignment of accreting matter: the
coherent alignment of the net spin and hence similar angular momentum
leads to disk dominated systems whereas the direct filamentary
accretion of gold gas is accompanied by spin misalignments and favors
the formation of spheroids.
 
Along these lines, here we study the relation between the tidal field
and spheroids with early massive black holes and the suppressed early
growth of black holes in regions of disk formation.  As an
illustration, we show that indeed where tidal fields are large ($t1 =
0.6$), a large scale filament forms (see
Figure~\ref{fig:BHhosts_disk}).  This induces significant tangential
motion in the accreting gas, leading to high angular momentum and the
formation of a massive early disk.  Conversely the tidal field
strength is low $t1=0.2$ in regions with thin converging filaments
which bring in cold material fast and lead to the formation of the
most massive black hole in a spheroidal dominated host. Tidal torque theory has
been shown by \citet{gonzalez16} to explain the initial spin of galaxies in
hydrodynamic simulations at high redshifts, and \cite{prieto15} have found that
halos with fewer filaments tend to have larger spin. This is in agreement
with the examples we have presented here, where the disc galaxies
tend to lie in regions with two main filaments, and the low angular momentum
galaxies that host black holes lie at the centers of a more symmetric 
arrangement of multiple filaments.

We now move on from these illustrative examples to investigate whether
these results also apply statistically to the galaxy morphology and
black holes accretion rates of the entire sample of massive galaxies
in BLueTides. We measure the tidal field strength for the 22156
galaxies with stellar masses $\gt 10^{8}\,{\rm M_{\odot}}$) that host
a black hole at $z=8$.  In Figure~\ref{fig:tidalfield} we plot the
mean $D/T$ ratio and Eddington accretion rate as a function of the
tidal field strength $t1$ (measured at a scale of 1 $\Mpc$)

We can see in the left panel of Figure~\ref{fig:tidalfield} that the
mean D/T ratio for the galaxies in \bluetides does show a strong
dependence on tidal field strength. As seen in the examples above, the
large tidal field environments induce coherent acquisition of angular
momentum and hence disk formation. Conversely, in the right panel of
Figure~\ref{fig:tidalfield} we find that on average the largest black
hole accretion rates are induced in regions with the smallest tidal
fields.  This is consistent with the fastest black hole growth at
these early times being associated to those region in which gas is
accreted cold and fast through direct filamentary accretion. The most
massive black holes thus may form even in the absence of mergers.

We quantify these relationships in two fashions. First, we compute the
Spearman rank correlation coefficient $r_{s}$ between $t$ and $D/T$
ratio and between $t$ and Eddington accretion rate.  We find $r_{s}
\sim 0.1$ and $r_s \sim - 0.05$, respectively with very high
significance (see the results in Table \ref{table:spearman}) This
indicates that although there is scatter about the relation, there is
an extremely significant correlation between $t$ and $D/T$ and an
anti-correlation between $t$ and accretion rate.  We have also fit a
simple linear relation between the values of $D/T$ averaged in bins of
$t$ and a linear fit to the Eddington accretion values in bins of $t$
also yields an acceptable. The parameters for these fits are also
given in Table \ref{table:spearman}.

These quantitative results confirm our earlier suggestion that regions
of significant tidal field strength hosting early massive galaxies
preferentially lead to the formation of disks, while the lowest tidal
fields lead on average to the more significant cold accretion and
black hole growth, and which is associated with the formation of
spheroid dominated systems.

\section{Conclusions}
We use the hydrodynamical cosmological simulation \bluetides to study
the origin of the most massive black holes at early times. Here we
have focussed our analysis on the massive galaxies and black holes at
$z\ge 8$ \citep[for earlier discussion of $z=6$ SDSS quasars, see
e.g.][]{DiMatteo2012, Feng2014}. In \citet{Feng2015} we discussed how
the most massive galaxies at these epochs are likely disk
dominated. Here we have shown that while the most massive disks can
still host massive black holes the most extreme early black hole
growth (reaching a $M_{\rm BH}$ a few $10^{8} \Msun$ at $z=8$) occurs
in spheroid-dominated extremely compact galaxies. The sites of
extremely rapid black hole growth hosted by spheroid-dominated
galaxies are the result of large-scale filamentary accretion of cold
gas from which the halo and eventually black holes can accrete new
material radially (at low angular momentum). The large scale, thin
filamentary structures surrounding these halo is a direct result of
the relatively low tidal field strength due to the surrounding
large-scale density field. In contrast, for halos located in strong
tidal fields, the surrounding filamentary structure is larger than the
halo allowing material to be accreted from different directions
producing more gradual accretion (deceleration along $t1$) parallel to
the large scale filament - and acceleration (along $t3$) -
perpendicular to the filaments, allowing material to accrete angular
momentum more coherently.

We find that star formation in the spheroidal hosts of massive BHs
prompts very rapid transformation of gas into stars and with BH
feedback heating and somewhat quenching the inner regions which result
in an older stellar populations. Disk dominated hosts tend to have
even higher star formation rates (less quenched by AGN feedback) and
which arise after a gradual ramp up in time as gas accretion is
somewhat delayed compared to the spheroids. We plan to explore the
observational signatures of these populations of stars in the most
massive disks versus the most massive black hole hosts in future work.

Our results suggest a new scenario for the origin of the most massive
early black holes which does not rely on major gas rich mergers. This
scenario offers clues to the origin of the most massive, rare high-z
black holes. These extreme black holes can be grown at the highest
rates in high density regions in the early Universe but not
necessarily the most extreme ones: what is important is that they grow
most rapidly in regions with relatively low tidal fields due to their
environments. The fact that the brightest quasars at these early times
host the most massive black holes, but do not necessarily have to live
in the densest regions of the universe, imply that they are unlikely
to be the precursors of the massive black holes in cluster of galaxies
today.  More likely their descendants will not be in privileged sites
but in rather isolated galaxies \citep[e.g.][]{Thomas2016} 

Our
  work  and the finding that  black hole growth is linked 
to tidal field theory offers
  some explanation as to the origin of the most massive BHs in most
  compact galaxies. It is interesting that \cite{Barber2016} using the
  EAGLE simulation at $z=0$ also found a strong link between
  compactness of the host and the most (positive) outliers in the
  $M_{\rm BH}-M_{*}$ relation.  These authors also find that the most
  extreme outliers grow rapidly in $M_{\rm BH}$ at early times to to lie
  well above the present day $M_{\rm BH} - M_{*}$ relation 
  and are consistent with undisturbed morphologies implied by
 late time observations. In the context of our findings it is likely
that indeed the
  most extreme massive black hole outliers are linked 
  to regions with lower tidal fields in the initial conditions.
While it is currently unfeasible (with any current HPC) to run
  \bluetides to $z=0$, we also plan to carry out a Dark-Matter Only
  \bluetides which will allow us to study the descendants of the the
  earliest supermassive BHs.

  The fastest black hole growth is then a direct consequence of the
  initial conditions of the density field.  Although the limited time
  evolution and somewhat limited numerical resolution of the large
  volumes of \bluetides precludes more detailed studies and
  investigations we plan to use these clues to re-simulate some of
  these systems at higher resolution \citep[][as we did with MB
  simulation, see e.g.]{Feng2014} with the goal of studying in more
  detail the dependence and constrain seed black hole scenarios.
    We note that \cite[][and references therein]{Danovich2015} have
    studied in detail the angular-momentum buildup in $high-z $ massive
    galaxies using high-resolution zoom cosmological simulations and
    have explicitly studied the link with early disc formation and
    tidal field theory.  In our work we illustrate that  black hole formation
    and angular momentum buildup can be understood within the same context.

\section{Acknowledgements}
We thank Y. Dubois and M. Volonteri for early discussion on the
importance of angular momentum build-up for early, massive BH
formation.  We acknowledge funding from NSF ACI-1036211, NSF
ACI-1614853, NSF AST-1517593, NSF AST-1009781, and the BlueWaters PAID
program. The \bluetides\ simulation was run on facilities on
BlueWaters at the National Center for Supercomputing Applications. SMW
acknowledges support from the UK Science and Technology Facilities
Council.

\bibliographystyle{mn2e}
\bibliography{main.bib}
\end{document}